\documentclass[eqsecnum,aps,showpacs,twocolumn]{revtex4}  
\usepackage{graphicx}

\begin{document}

\title{Effect of an impulsive force on vortices in a rotating Bose-Einstein
condensate}

\author{Sadhan K. Adhikari}

\author{Paulsamy Muruganandam}

\affiliation{Instituto de F\'{\i}sica Te\'orica, Universidade Estadual
Paulista, 01.405-900 S\~ao Paulo, S\~ao Paulo, Brazil}

\date{\today}

\begin{abstract}

The effects of a sudden increase and decrease of the interatomic interaction
and harmonic-oscillator trapping potential on  vortices in a quasi
two-dimensional rotating Bose-Einstein condensate are investigated using the
mean-field Gross-Pitaevskii equation.  Upon increasing the strength of
interaction suddenly the condensate enters a nonstationary oscillating phase
which starts to develop more vortices. The opposite happens if the strength is
reduced suddenly. Eventually, the number of vortices attains a final value at
large times. Similarly, the number of vortices increases (decreases) upon a
sudden reduction (augmentation) in the trapping potential. We also study the
decay of vortices when the rotation of the condensate is suddenly stopped. 
Upon a free expansion of a rotating BEC with vortices the radius of the vortex
core increases more rapidly than the radius of the condensate. This makes the
counting and detection of multiple vortex easier after a free expansion.

\end{abstract}


\maketitle

\section{Introduction}
 
Since the successful detection \cite{1} of Bose-Einstein condensates (BEC)
in dilute trapped bosonic atoms at ultra-low temperature, one problem of
extreme interest is the formation of vortices in a rotating condensate
in an axially symmetric
trap. 
The experimental observation
\cite{exp,exp1,expm,jila,hal,hal1,mit,mit1} of vortices in
the condensate has intensified theoretical investigations on vortices
using the mean-field Gross-Pitaevskii (GP) equation \cite{gp}. For large
frequencies of rotation, a very large number of vortices arranged in a
triangular lattice, frequently referred to as a vortex lattice, have been
observed.  The
 vortex lattices are highly excited collective states of BECs with an
angular momentum of up to $60\hbar$  per particle \cite{mit}. 
There have
been many theoretical studies on different aspects of BEC \cite{2,3,11}
and specially, on the vortices in a rotating BEC in 
axially symmetric traps
\cite{2a,2c,2d,2e,sin,2f,2ff,2f1,2g,2h,fc,ad,ueda,dis}.

When an ordinary fluid is placed in a rotating container the fluid rotates
with the container because of viscous force. For superfluid $^4$He(II) in
a rotating container no motion of the fluid is observed below a critical
rotational frequency.  Above the critical frequency, lines of singularity
appear in its velocity field.  These lines of singularity correspond to
vortices and appear in the form of a lattice specially for large
rotational frequencies.  Superfluidity exhibits many peculiar features:
absence of viscosity, reduction of moment of inertia, occurrence of
persistent current, appearance of quantized vortices among others
\cite{11}. The formation of vortex lattice in a rotating $^4$He(II)  is
intimately related to its superfluidity and is a manifestation of
quantized vortices \cite{exp,exp1,expm,jila,mit,mit1}. However, because of
the
strong interaction between $^4$He atoms, the theoretical description of
this system is not an easy task.

Similar vortices can be generated in theoretical mean-field models of
trapped BEC based on the GP equations
\cite{2a,2c,2d,2e,sin,2f,2ff,2f1,2g,2h,fc,ad,ueda} and have been observed
experimentally \cite{exp,exp1,expm,jila,hal,hal1,mit,mit1} in BEC with
repulsive interaction. In contrast to liquid $^4$He(II), a trapped BEC is
a very dilute and weakly interacting system, which makes a mean-field
analysis appropriate. This analogy of a trapped BEC with liquid $^4$He(II)
suggests the presence of superfluidity in the BEC.  Many different ways
for generating quantized vortices in a BEC have been suggested \cite{2d} ,
e.g., via spontaneous vortex formation in evaporative cooling \cite{2f},
via controlled excitation to an excited state of atoms \cite{2ff}, by
stirring a BEC using an external laser with an angular frequency above a
critical value for generating a quantized vortex with a single quantum
\cite{2c}, or by the rotation of an axially symmetric trap with an angular
frequency above a similar critical value \cite{2g}.

Experimentally, quantized vortices have been generated by different groups
in a rotating BEC using a variety of techniques. Vortices have been
detected in a $^{87}$Rb condensate in a cylindrical trap by Madison et al
at ENS \cite{exp}, where angular momentum is generated by a stirring laser
beam.  At MIT Raman et al \cite{expm} also studied the nucleation of
vortices in a Bose-Einstein condensate stirred by a laser beam.  They have
been observed by Matthews et al \cite{jila} at JILA in coupled BEC's
comprised of two spin states of $^{87}$Rb in a spherical trap, where
angular momentum is generated by a controlled excitation of the atoms
between the two states. The tilting of the vortex axis in a nonspherical
trapping potential has been studied at JILA by Haljan et al \cite{hal}.  
Vortices have also been generated by evaporatively spinning up a normal
gas of $^{87}$Rb and then cooling below quantum degeneracy at JILA by
Haljan et al \cite{hal1}. A large number of vortices in the form of a
lattice have been observed at MIT by Abo-Shaeer et al \cite{expm,mit} as
well as at ENS by Madison et al \cite{exp} for sufficiently large
rotational frequency. The interesting dynamics of the formation and decay
of vortices has also been studied \cite{mit1}. Madison et al \cite{exp}
also confirmed that before a vortex pattern is formed the condensate
passes through dynamically unstable distorted configurations. The
dynamical instability transforms the distorted condensate to an
appropriately symmetric condensate with vortices \cite{sin,ueda}.

There have been theoretical studies of how vortices are generated in a BEC
via these dynamically unstable configurations using the mean-field GP
equations in two \cite{ueda} and three \cite{fc} dimensions.  The
theoretical study of the dynamical instability using mean-field approach
in the generation of vortex is of interest as one can then compare the
results of the models with experiments. This will provide a more stringent
test of the mean-field theories than the comparison of stable
configurations of BEC.

However, there is another type of instability which has been studied in
relation to a nonrotating BEC under the application of an impulsive
force.  One can now
suddenly change the trapping potential on a stable BEC. This can be done
in experiment by varying the current in the  coil responsible for
confinement.  One can also
change the strength of interatomic interaction suddenly by changing the
magnetic field near a Feshbach resonance \cite{yy1}. These set in
dynamical instabilities in the BEC, which can be studied both
experimentally \cite{ex} and theoretically \cite{th}. 

Similar impulsive force can be applied on a rotating BEC with vortices  and the
resultant oscillation studied.  One can study how the number of vortices change
after the application of the impulsive force on a rotating BEC.  In this paper
we perform such  mean-field analyses of the dynamics of  a rotating BEC subject
to three  types of impulsive forces. We consider  a sudden change in (i) the
strength of interaction near a Feshbach resonance \cite{yy1} induced by a
variation of an external magnetic field,  (ii) the trapping potential, and
(iii) the rotation of the BEC. Specifically, we study the dynamics  when the 
strength of interaction or  the trapping potential is suddenly increased or
reduced.   We also study the decay of vortices once the rotation of the
condensate is suddenly stopped. There have been certain experimental
measurements on such decay of vortices \cite{exp,mit1}. Finally, the trapping
potential can be removed suddenly  in a rotating BEC with vortices  and the
resultant expansion studied.

We base our study on the numerical solution \cite{3,2a,2h} of the nonlinear
time-dependent mean-field GP  equation \cite{gp}, which should provide a
faithful description of the formation and evolution of vortices. The
experiments on the vortices   used a highly anisotropic trap which may be
simulated by a quasi two-dimensional model. Hence as in the recent studies on
the subject in Refs. \cite{ad,ueda} we employ the GP equation in two
dimensions. We use a split-step method where in the time evolution of the GP
equation the kinetic energy and angular momentum (derivative) terms in the two
directions are dealt with in two independent steps by the   Crank-Nicholson 
rule  complimented by the known boundary conditions \cite{2h,koo}. All the
harmonic oscillator and the nonlinear terms are treated in a third separate
step.  We find that this time-dependent split-step  approach leads to good
convergence.

In Sec. II we describe briefly the time-dependent  GP equation for a rotating
BEC and a numerical method for its solution.  In Sec. III we report the
numerical results of the present investigation  and finally, in Sec. IV we give
a summary of our study. 

\section{Nonlinear Gross-Pitaevskii Equation}

At zero temperature, the time-dependent Bose-Einstein condensate wave
function $\Psi({\bf r};\tau)$ in two dimensions at position ${\bf r}$ and
time $\tau $ may
be described by the following  mean-field nonlinear GP equation
\cite{gp,11}
\begin{eqnarray}\label{a} \biggr[ -\frac{\hbar^2}{2m}\nabla^2
&+& V({\bf r})  
+ gN|\Psi({\bf
r};\tau)|^2- \hat \Omega L_{\hat z} \nonumber \\
&-&i\hbar\frac{\partial
}{\partial \tau} \biggr]\Psi({\bf r};\tau)=0.   \end{eqnarray} Here $m$
is
the mass and  $N$ the number of atoms in the
condensate, 
 $g$ the strength of interatomic interaction. 
The trapping potential in two dimensions  may be written as  $  V({\bf
r}) =\frac{1}{2}m \omega ^2(\hat x^2+ \hat y^2)$ where 
$\omega$ is the angular frequency. The rotational term
 $-\hat \Omega L_{\hat z}
= i\hbar \hat \Omega (\hat x\partial _{\hat y} -\hat  y\partial_{\hat x})$
where $\hat \Omega $ is the
rotational frequency along the $\hat z$ direction with $L_{\hat z}$ the
$\hat z$ component
of angular momentum.  
The normalization condition of the wave
function is
$ \int d{\bf r} |\Psi({\bf r};\tau)|^2 = 1. $
The GP
equation (\ref{a}) can 
accommodate
quantized vortex
states with rotational motion of the condensate around the $\hat z$ axis.

It is convenient to use the dimensionless variables   
defined by $x =\hat x  /l$,  $y=\hat y/l$,   $t=\tau \omega/2, $
$l\equiv \sqrt {\hbar/(m\omega)}$, $\Omega=2\hat \Omega/\omega $, 
and  
$\varphi(x,y;t)=l 
{ \Psi({\bf r};t)}$. In these units Eq. (\ref{a}) becomes 
\begin{eqnarray}\label{d}
\biggr[ -\frac{\partial^2}{\partial
x^2}&-&\frac{\partial^2}{\partial
y^2}
+\left(x^2+y^2\right)+i\Omega\left( x\frac{\partial }{\partial y}
- y\frac{\partial }{\partial x}\right)   
\nonumber \\
&+& {\cal N}\left| {\varphi({x,y};t)}\right|^2 
-i\frac{\partial
}{\partial t} \biggr]\varphi({ x,y};t)=0, 
\end{eqnarray}
where
$ {\cal N} =  2g Nm  /\hbar^2.$ 
The normalization condition  of the wave
function is \begin{equation}\label{5}  \int_{-\infty} ^\infty
dx \int _{-\infty}^\infty dy|\varphi(x,y;t)|
^2 =1.  \end{equation}

The explicit inclusion of the rotational energy in the GP hamiltonian
above simulates the time evolution of a rotational BEC which permits
multiple-vortex formation. The initial input solution to the GP equation
is a circularly symmetric state obtained with $\Omega =0$. The method of
solution with $\Omega \ne 0$ is time iteration using an appropriate
algorithm.  If one iterates the GP equation above in time using the full
angular momentum term the circular symmetry is maintained always and a
solution with multiple vortices is not generated. It is essential that the
circular symmetry of the solution is broken at some stage before multiple
vortices are generated during time iteration.

The circular symmetry of the
solution can be broken in the process of time evolution of
the GP equation above by
introducing  \cite{fc} a phase in the initial wave function for $\Omega
=0$
of the form
\begin{equation}\label{ph} \delta= \sum_{i} \arctan [(y-y_i)/(x-x_i)],
\end{equation} where $x_i$ and $y_i$ are a set of parameters such that the
positions $(x_i,y_i)$ fall inside the boundary of the BEC. This
initial phase preserves the norm of the wave function but accelerates the
formation of vortex nucleation after time evolution of the GP equation.
However, there is no direct relation between the parameters $(x_i,y_i)$
and the positions or the numbers of the vortices in the final BEC. Another
way of facilitating vortex formation is to include a small absorptive term
in the GP equation as has been advocated in Ref. \cite{2f} and employed in
Ref. \cite{ueda}.

We solve the GP equation (\ref{d}) with  $\Omega = 0 $  
using a
time-iteration method elaborated in Refs. \cite{2h,koo} starting
from the analytic two-dimensional harmonic-oscillator solution for ${\cal N}=0$.  The full
GP Hamiltonian is conveniently broken into three parts $H_x$, $H_y$ and
$H_{\mbox{res}}$ $-$ the first containing the $x$-dependent derivative
terms, the second containing the $y$-dependent derivative terms with
the harmonic-oscillator and nonlinear interaction terms contributing to
the third residual part. The GP equations along the $x$ and $y$ directions
are discretized
on a two-dimensional set of grid points $N_x \times N_y$ using the
Crank-Nicholson discretization method \cite{koo}. 
The second (first) derivatives are discretized using a three-point (two-point)
finite difference rule.
Effectively, each time iteration of the GP equation is broken up into
three parts $-$ using $H_x$, $H_y$ and $H_{\mbox{res}}$. 
The first two parts lead to two sets of  tridiagonal equations
along $x$ and $y$ directions which are solved alternately by the Gaussian
elimination method  \cite{koo}.  The third part does not contain any
derivative and is solved essentially exactly. For a small time
step $\Delta$ the error involved in this break-up procedure along $x$ and
$y$ directions is quadratic in $\Delta$ and hence can be neglected.
Using the above procedure the solution of the GP equation corresponding to
$\Omega
=0$ and ${\cal N}\ne 0$
is obtained. This solution serves as the initial input  for the
simulation of multiple vortex generation  using the time iteration of the
GP equation.

The initial solution with $\Omega =0$ is circularly symmetric. For the
generation of multiple vortex the circular symmetry is broken by
introducing the phase $\delta$ of Eq. (\ref{ph}) to the solution with
$\Omega =0$ and iterating the GP equation in time with full $\Omega$. We
also included a small dissipative term in the GP equation.  As the
formation and decay of vortices is intrinsically dissipative \cite{mit1},
it is more realistic to include a small dissipative term \cite{ueda,dis}.
However, from a mathematical point of view, vortices and vortex lattice
can
be generated in numerical simulation without including any  dissipative
term \cite{fc}.  The time evolution of the GP equation is performed as
described
above using three independent steps. The system passes through dynamically
unstable deformed configurations and during time iteration the multiple
vortex centers appear for $\Omega$ greater than a critical value.

\section{Numerical Results}

In the present numerical study on vortices we use an equilibrated
condensate with ${\cal N}=500$ trapped in the stationary potential as the
initial state. Figure 1 shows the dynamics of the condensate as it starts
to rotate suddenly with an angular frequency $\Omega =1.5$. In the
simulation we use as input the wave function with phase $\delta$ given by
Eq. (\ref{ph}). In the beginning the system passes through deformed
configurations with unstable oscillating boundaries. The vortices are
generated in the peripheral region via the oscillation of the boundary.
More and more vortices are generated which move towards the center and
which arrange themselves in a regular form in the condensate.  For a
sufficiently large number of vortices they are arranged as in a lattice
and maintain that pattern during rotation. When such a regular
equilibrated final pattern is formed the condensate attains the symmetry
of the problem $-$ circular in the present case. 
\begin{figure}[!ht]
\begin{center}
\includegraphics[width=\linewidth]{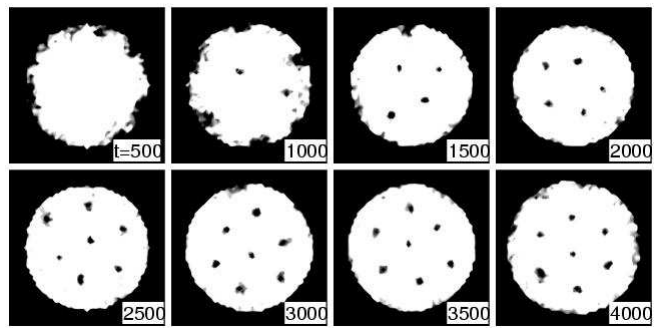}
\end{center}
\caption{Formation of the vortices for ${\cal N} = 500$ and $\Omega =1.5$ at
different times; size of each square is $10\times 10$.}
\end{figure}

In actual experiment the size of the condensate reduces  with time due to
inelastic atomic collisions \cite{mit}. However, throughout the present study
we renormalize  the  number of particles during time iteration so that the
size  is conserved.  This should not affect the generation of vortices. 

For $\Omega =1.5$ and ${\cal N}=500$ one has a hexagonal lattice structure with
seven vortices at large times.  In recent theoretical studies \cite{fc,ad,ueda}
some general features of vortex formation have been established which are
confirmed in the present study. For a fixed large ${\cal N}$ there is a
critical $\Omega $ for the appearance of a certain number of vortices
\cite{exp,mit}. The critical value of $\Omega $ increases as the number of
vortices increases. For a fixed $\Omega $ the number of vortices increases as
${\cal N}$ increases. It is interesting to recall that for ${\cal N}=0$ only
states with a single vortex at the center can be formed.

An interesting set of experiments can be performed where on a stable rotating
BEC with vortices  the strength of atomic interaction  or the harmonic
oscillator trapping potential is  suddenly changed.  Once either of these
parameters are changed suddenly on a vortex lattice with hexagonal structure,
the system enters again a   dynamically unstable asymmetric  configuration with
highly oscillating deformed boundary. Through the oscillating deformed boundary
new vortices can enter the condensate or some of the existing vortices can get
out. Eventually, a new equilibrated BEC with vortices  is formed in the
condensate with circular symmetry.  However, as the condensate  oscillates
after the application of the impulsive force it is more difficult to obtain
perfect symmetry numerically  in the final state.  Small imperfections remain
near the boundary and in the arrangement of  the vortices in the final state. 

\begin{figure}[!ht]

\begin{center}
\includegraphics[width=\linewidth]{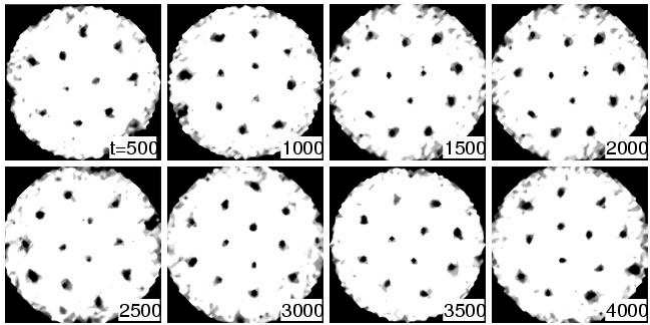}
\end{center}

\caption{ Dynamics of the condensate of Fig. 1 at different times $t$ upon
sudden increase of $\cal N$ from 500 to 1200 at $t=0$; size of each square is
$10\times 10$.}

\end{figure}
 
\begin{figure}[!ht]
 
\begin{center}
\includegraphics[width=\linewidth]{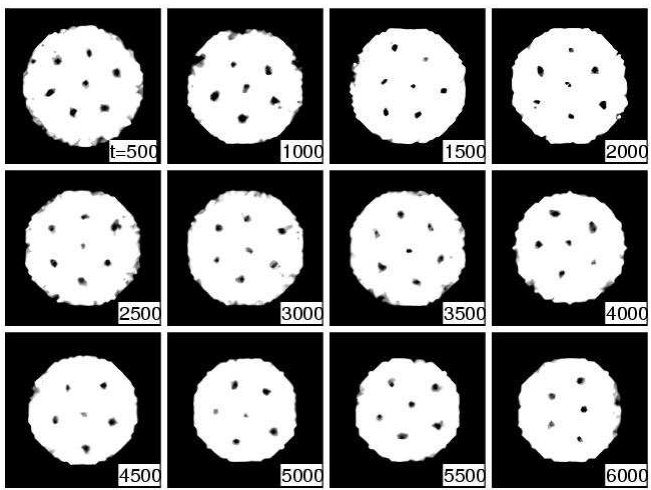}
\end{center}
\caption{Dynamics of the condensate of Fig. 1 at different times $t$ upon
sudden decrease of $\cal N$ from 500 to 100 at $t=0$; size of each square is
$10\times 10$. }
\end{figure}

In the simulation presented in Fig. 2 the nonlinear parameter $\cal N$ on
the hexagonal vortex lattice of Fig. 1 is suddenly changed to 1200 from
500 at $t=0$ and the evolution of the resultant condensate studied. This
corresponds to an increase of the strength of interatomic interaction by a
factor 2.4. The size of the condensate increases and it accommodates more
vortices. The system seems to keep on oscillating with $(12\pm 1)$
vortices.  This oscillation seems to be a natural consequence of the
sudden change of the strength of interaction. Next we consider in Fig. 3
the sudden change in the nonlinear parameter $\cal N$ from 500 to 100 at
$t=0$ on the final configuration  of Fig. 1. With the reduction of the
strength of interaction by a factor 0.2, the condensate shrinks in size by
passing through dynamically unstable and asymmetric configurations and
starts to oscillate.  The condensate tries to reach a equilibrated stage
by getting rid of vortices. However, at large times the system continues
to oscillate with ($6\pm 1$) vortices.
\begin{figure}[!ht]
\begin{center}
\includegraphics[width=\linewidth]{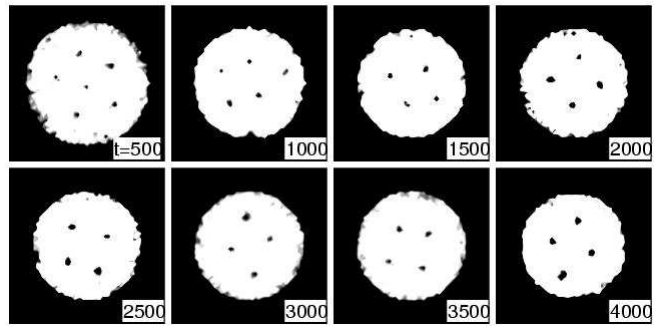}
\end{center}
\caption{ Dynamics of the condensate of Fig. 1 at different times upon sudden
increase of the harmonic oscillator trapping potential by a factor of 1.5 at
$t=0$; size of each square is  $10\times 10$.}
\end{figure}
\begin{figure}[!ht]
\begin{center}
\includegraphics[width=\linewidth]{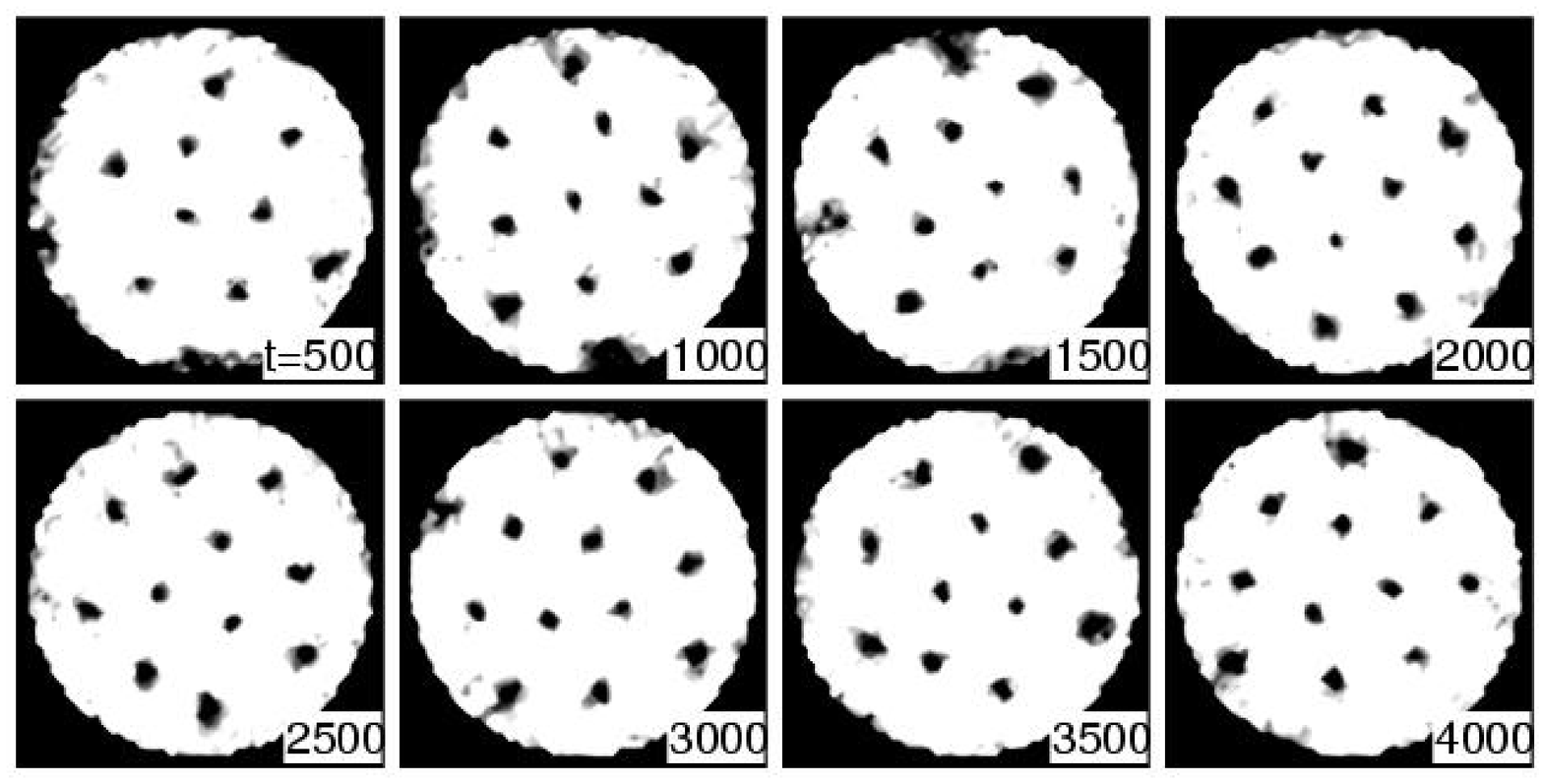}
\end{center}
\caption{ Dynamics of the condensate of Fig. 1 at different times upon sudden
decrease of the harmonic oscillator trapping potential by a factor of 0.75  at
$t=0$; size of each square is $10\times 10$. }
\end{figure}
\begin{figure}[!ht] 
\begin{center}
\includegraphics[width=\linewidth]{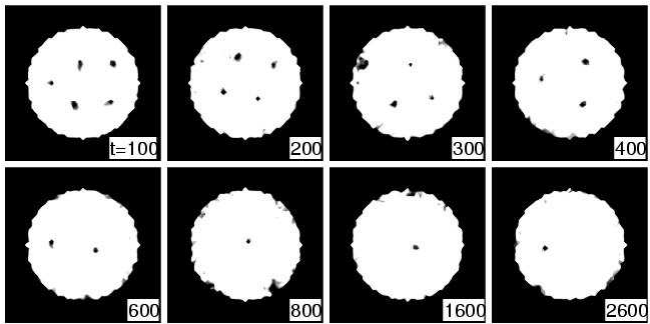}
\end{center}
\caption{ Decay of the condensate of Fig. 1 at different times after the
rotation is suddenly stopped at $t=0$; size of each square is $10\times 10$.}
\end{figure}

Next we consider a sudden change in the harmonic oscillator potential.
First, the harmonic oscillator potential is increased by a factor of 1.5
on the rotating BEC of Fig. 1 at $t=0$. The time evolution of the
resultant condensate is shown in Fig. 4. With the sudden increase in the
trapping potential the condensate starts to shrink and oscillate. It
passes through dynamically unstable asymmetric configurations and reaches
an equilibrium stage by getting rid of a couple of vortices. Eventually, a
condensate emerges with four vortices. The system executes small
oscillations at large times with a fixed number of vortices. In the next
simulation the harmonic oscillator potential is reduced by a factor of
0.75 on the configuration  of Fig. 1 at $t=0$. Because of the sudden
decrease in the trapping potential the condensate starts to swell and
oscillate. More vortices are generated near the boundary which move
inwards. Eventually, one has a larger condensate with 11 vortices at large
times.

We study the decay of the vortices  of Fig. 1 once the rotation of the BEC is
suddenly stopped at $t=0$. The dynamics of the decay of vortices has been
studied experimentally in Refs. \cite{exp,mit1}. The numerical simulation of
the decay of vortices  is shown in Fig. 6. The vortices do not decay
immediately after stopping the rotation. Rather they stay for a long time after
the rotation stops. As time passes vortices get out of the condensate until one
with a single  vortex results. This BEC with a single vortex survives for a
long time. If we consider $\omega =2\pi\times 219 $ s$^{-1}$ as in the
experiment  of Madison et al \cite{exp}, one unit of present dimensionless
time  is 0.000726 s. The system gets rid of six vortices in about 600 units of
time or in 0.4 s. However, the state with one vortex stays up to $t\approx
3000$, which gives a lifetime of about 2400 units of time or about 1.6 s. 
Madison et al \cite{exp} measured the half life of a single vortex state and
found it to be about 1 s, which is of the same order of magnitude as in the
present simulation. However, a quantitative comparison between the two is not
to the point because of approximate nature of the present simulation compared
to experiment. The present calculation used a quasi two-dimensional model
opposed to a full three-dimensional one. Also, the nonlinearity of the present
model is not the same as in the experiment. However,  in a rotating  BEC  with
many vortices, most of the vortices decay within one second whereas a small
number of vortices survive for a long time.  This was also experimentally
confirmed by Abo-Shaeer et al \cite{mit1} who used $\omega \approx  2\pi \times
85$ s$^{-1}$.
\begin{figure}[!ht]
\begin{center}
\includegraphics[width=\linewidth]{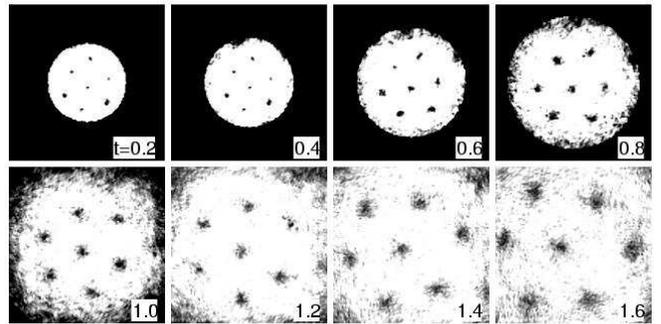}
\end{center}
\caption{  Free expansion of the condensate Fig. 1 at different times after 
the removal of the trap at $t=0$; size of each square is $20\times 20$.  }
\end{figure}

Finally, we consider the free expansion of the vortices in Fig. 1 when the
harmonic oscillator trap is suddenly removed. The dynamics of free expansion is
shown in Fig. 7. It is interesting that the vortex pattern is not destroyed
immediately after the removal of the trap. On the other hand, upon expansion
the radius of the vortex core is found to increase more rapidly than the radius
of the condensate.  Due to the very small radius of the vortex core it is
difficult to observe the vortices and count them properly in an experiment
\cite{mit1}. However, because of the increase in the radius of the vortex core
after a free expansion the detection and counting of vortices will be easier
from a picture of the condensate after expansion \cite{mit1}. In Fig. 7 the
vortices are more prominently visible at $t=1.6$ than at $t=0.2$. Using the
radial trapping frequency $\omega = 2\pi \times 85$ s$^{-1}$ as in the
experiment of Abo-Shaeer et al \cite{mit1}, $t=1.6$ of Fig. 7 corresponds to
about 3 ms. In the actual experiment \cite{mit1} after 42.5 ms of expansion the
vortex cores were magnified 20 times their size in the trap. The present
simulation has again led to the right order of magnitude. 

\section{Summary}

We have investigated the effect of an impulsive force on a rotating BEC with
vortices using the mean-field GP equation. We have considered the effect of
suddenly increasing or decreasing the strength of atomic interaction and the
harmonic oscillator trapping potential. Upon application of such an impulsive
force the condensate passes through highly unstable distorted configurations
with oscillating boundary. New vortices can enter the condensate via the
oscillating boundary. Also the condensate can get rid of some of the vortices
in a similar fashion. Eventually, a (more) symmetric condensate with a
different number of vortices result. When the atomic interaction is suddenly
increased (decreased) the final condensate grows (shrinks) in size and
accommodates more (less) vortices.  When the trapping potential  is suddenly
increased (decreased) the final condensate shrinks (grows) in size and
accommodates less (more)  vortices.  Experiments can be performed when similar
impulsive force in applied on a rotating BEC with vortices  and compared with
the mean-field prediction.

We also study the decay of vortices in a rotating BEC when the rotation is
suddenly stopped. Most of the vortices decay quickly (in a fraction of a 
second), but a few vortices  survive during a relatively long time before they
decay. In a typical experimental situation \cite{exp,mit1} the single vortex at
the end survives for a unusually long interval of time (a couple of seconds).

In another simulation of a free expansion of a rotating BEC  we find that  the
vortices are not immediately lost during expansion.  Rather, upon expansion the
radius of the vortex core increases at a faster rate than that of the
condensate. Hence the vortices are more prominently visible in a picture of the
condensate during the free expansion of the condensate.   Experimentally, it is
difficult to observe and count the vortices  correctly as the vortex cores have
usually  small radii \cite{mit1}. The present simulation suggests that the free
expansion could aid in the observation and counting of vortices in a rotating
BEC, specially, when there is a large number of vortices. This was noted in the
experiment of Abo-Shaeer et al \cite{mit1}. In a typical experimental
situation, on free expansion the vortices may be observed during few  tens of 
milliseconds.

\acknowledgments 

The work is supported in part by the Conselho Nacional de Desenvolvimento
Cient\'\i fico e Tecnol\'ogico and Funda\c c\~ao de Amparo \`a Pesquisa do
Estado de S\~ao Paulo of Brazil.

\end{document}